\documentclass[letter]{aa} 
\usepackage{graphicx}
\usepackage{epstopdf}
\usepackage{txfonts}
\usepackage{natbib} 
\bibpunct{(}{)}{;}{a}{}{,} 
\newcommand{\caii}{\ion{Ca}{II}}
\begin{document}
\title{Small-scale swirl events in the quiet Sun chromosphere} 

\author{Sven Wedemeyer-B\"ohm\inst{1,2}\thanks{Marie Curie Intra-European Fellow of the European Commission}
\and 
Luc Rouppe van der Voort\inst{1}
}

\offprints{sven.wedemeyer-bohm@astro.uio.no}

\institute{Institute of Theoretical Astrophysics, University of Oslo,
  P.O. Box 1029 Blindern, N-0315 Oslo, Norway
  \and
  Center of Mathematics for Applications, University of Oslo,
  Box 1053 Blindern, N-0316 Oslo, Norway
}

\date{Received 30 September 2009; accepted 14 October 2009}

\abstract{Recent progress in instrumentation enables solar observations with high 
resolution simultaneously in the spatial, temporal, and spectral domains.
} 
{We use such high-resolution observations to study small-scale structures and dynamics 
in the chromosphere of the quiet Sun. 
} 
{We analyse time series of spectral scans through the \ion{Ca}{II}~854.2~nm spectral line 
obtained with the CRISP instrument at the Swedish 1-m Solar Telescope. The targets are quiet 
Sun regions inside coronal holes close to disc-centre. 
} 
{The line core maps exhibit relatively few fibrils compared to what is normally observed in 
quiet Sun regions outside coronal holes.
The time series show a chaotic and dynamic scene that includes spatially confined 
``swirl'' events.
These events feature dark and bright rotating patches, which can consist of arcs, 
spiral arms, rings or ring fragments. 
The width of the fragments typically appears to be of the order of only 0\,\farcs2, 
which is close to the effective spatial resolution. 
They exhibit Doppler shifts of $-2$ to $-4$\,km$/$s but sometimes up to $-7$\,km$/$s, 
indicating fast upflows. 
The diameter of a swirl is usually of the order of 2\,\arcsec.
At the location of these swirls, the line wing and wide-band maps show close groups 
of photospheric bright points that move with respect to each other.
}  
{A likely explanation is that the relative motion of the bright points twists the 
associated magnetic field in the chromosphere above.  
Plasma or propagating waves may then spiral upwards guided by 
the magnetic flux structure, thereby producing the observed intensity signature of 
Doppler-shifted ring fragments. 
}


\maketitle
%
\section{Introduction}

Advances in observational performance have revolutionised our understanding of 
the solar chromosphere over the last years.  
It is in particular the combination of high spectral, temporal, and spatial 
resolution that let us discover details and phenomena hitherto unaccessible. 
Our picture of the solar atmosphere changed from a static plane-parallel stratification 
to a very complex compound of intricately coupled dynamic domains 
(see recent reviews by e.g. Schrijver 2001; Judge 2006; Rutten 2006,
2007;Wedemeyer-B{\"o}hm et al. 2009). 
And yet many details concerning the chromosphere and its connection 
to the layers above and below remain elusive.  
The progress in our understanding is certainly hampered by the complexity and 
accessibility of currently available diagnostics probing that atmospheric layer. 
Among them are the spectral lines of \ion{Ca}{II}, which are 
formed over a very extended height range in the atmosphere. 
Only the line cores truly originate from above the photosphere.  
Observing the line core exclusively therefore requires filters with a very narrow 
transmitted wavelength range, as the width of the line cores is only of the order 
of 100\,pm or less. 

Considering the spectral, temporal, and spatial domains, instrumental limitations 
enforce compromises which until recently only allowed high resolution in one or two 
of the domains, at the cost of the remaining one(s). 
This situation has now improved substantially with the installation of fast 
two-dimensional spectropolarimetric imagers at solar telescopes with large 
aperture. 
Examples are the IBIS instrument
(Cavallini 2006) at
the Dunn Solar Telescope, CRISP (Scharmer et al. 2008) at
the Swedish Solar Telescope (SST), 
and the G{\"o}ttingen Fabry-P{\'e}rot (Puschmann et al. 2006)
Achieving high resolution simultaneously in all these three domains is like opening 
a new observing window on the chromosphere. 
Here we report on the discovery of small but fast rotating swirls in the chromosphere. 
There are quite a few examples known of rotating or vortex-like motions on the Sun: 
on large scales in the form of 
rotating sunspots (e.g., Brown et al. 2003), 
on smaller scales like vortices in the photospheric granulation 
(Brandt et al. 1988),
and on the smallest scales in the form of whirlpool motion by magnetic bright points
in the inter-granular lanes 
(Bonet et al. 2008).
This rotating motion receives considerable interest since these photospheric 
motions have a profound effect on the outer-atmospheric magnetic fields as they 
have their roots in the photoshere. 
The stresses that are built up through rotation of magnetic fields are linked to 
the onset of solar flares and coronal heating 
(e.g., Parker 1983).

After this introduction we describe the observations in Sect.~\ref{sec:observ}, 
which are analysed in Sect.~\ref{sec:ana}.  
A discussion and conclusions are given in Sect.~\ref{sec:discus}. 

\section{Observations}
\label{sec:observ}

\begin{figure*}[ht]
\centering 
\includegraphics{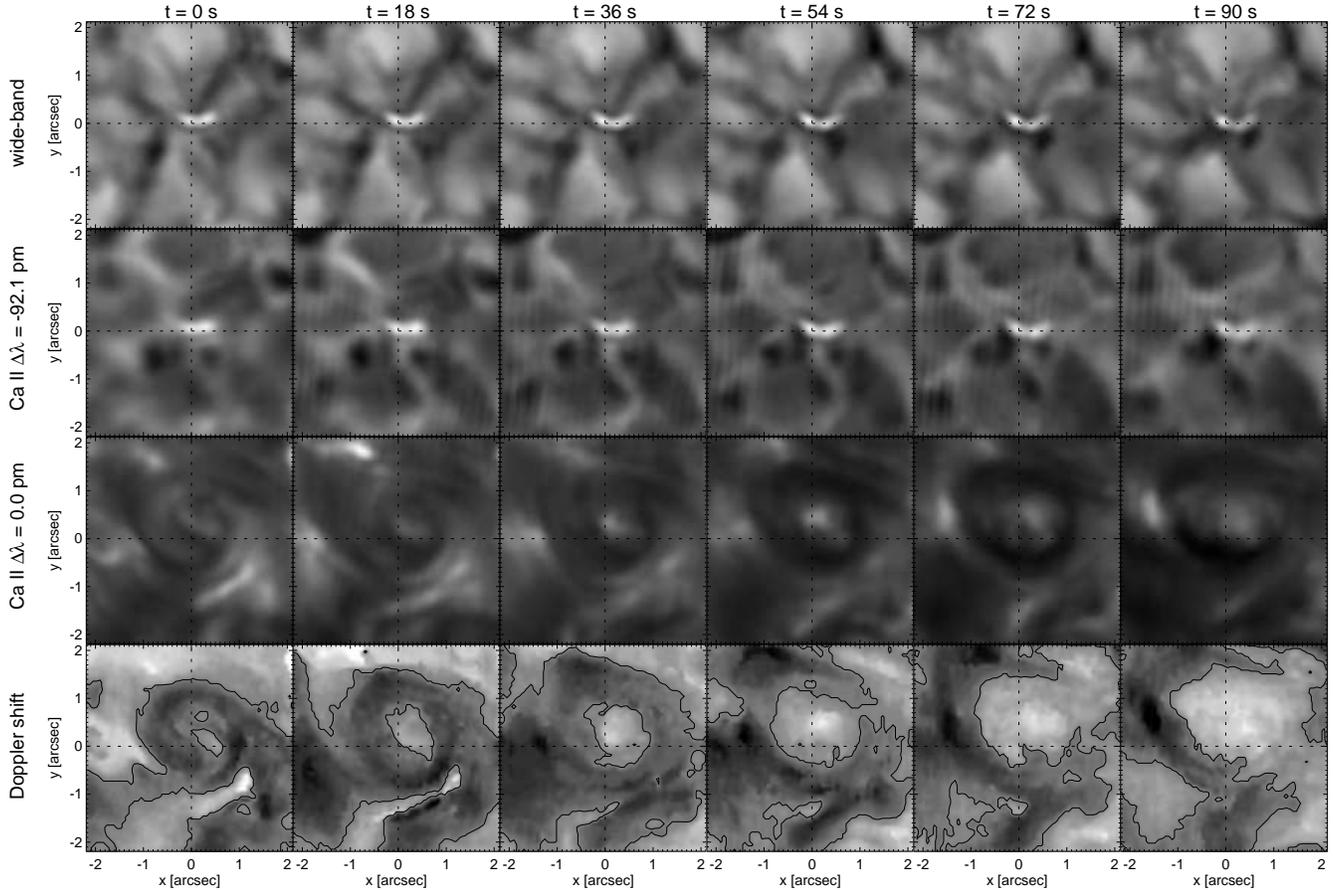}
\caption{Temporal evolution of a swirl event as seen in close-ups of intensity maps 
in the wide-band (top row), Ca line wing (upper middle), 
Ca line core (lower middle), and Doppler shift (bottom).  
The columns from left to right show every other time step ($\Delta t = 18$\,s). 
The black contours in the bottom row mark a zero Doppler shift. The grey scale 
of the Doppler shift is from -5.5 to $+5.5$\,km/s
with negative values corresponding to blueshifts and thus upflows. 
The temporal evolution is presented in a movie provided as online material.
}
\label{fig:swirl}
\end{figure*}

The observations were obtained with the CRisp Imaging
SpectroPolarimeter (CRISP, Scharmer et al. 2008) 
installed at the Swedish 1-m Solar Telescope
(SST, Scharmer et al. 2003) 
on La Palma (Spain).
CRISP is a spectropolarimeter that includes a dual Fabry-P{\'e}rot
interferometer and is capable of fast wavelength tuning
($\lesssim$50~ms), which makes it ideally suited for imaging the dynamic
chromosphere.
We observed the \ion{Ca}{II}~854.2~nm line (filter transmission FWHM
11.1~pm) and analyse two time sequences from two days in 2008,
targeting equatorial coronal holes close to disc center 
(field of view, FOV\,$\approx\,71\,\arcsec\,\times\,71\,\arcsec$). 
All data sets are complemented with wide-band images from the CRISP prefilter
(FWHM\,$=\,0.93$\,nm centered on 854.2~nm).  
On June 15th 2008, a series of H$\alpha$ images was recorded directly after the Ca 
scan at the same position. 
More details on the data are presented in Table~\ref{tab:datasets}.
With the use of adaptive optics combined with Multi-Object Multi-Frame
Blind Deconvolution
(MOMFBD, van Noort et al. 2005) 
image restoration, the time sequences are of excellent quality,
achieving a spatial resolution close to the diffraction limit
($\lambda/D = 0\,\farcs18$ at 854.2~nm) for the majority of the images.
%
\begin{table}[b]
\begin{minipage}[t]{\columnwidth}
\label{tab:datasets}
\caption{\caii\,854.2 line scan sequences from June 2008.\vspace*{-3mm}}
\begin{center}
\renewcommand{\footnoterule}{}  
\begin{tabular}{ccccccc}
\hline
data set&
\hspace*{-1mm}$\mu$&
$\lambda$\hspace*{-1mm}&
\hspace*{-1mm}$(\Delta \lambda)_\mathrm{core}$\footnote{wavelength sampling for $-70$\,pm\,$<\,\lambda\,<\,70$\,pm (core)}\hspace*{-1mm}&
\hspace*{-1mm}$(\Delta \lambda)_\mathrm{wing}$\footnote{wavelength sampling for $\lambda\,>\,70$\,pm, $\lambda\,<\,-70$\,pm (wings)}&
$\Delta{t}$& 
dur.\\
(date)&&[pm]&[pm]&[pm]&[s]&[min]\\
\hline
1\,(13th)&\hspace*{-1mm}0.92&  -92.0 -- +19.4  & 4.8&  4.8&  9& 33\\
2\,(15th)&\hspace*{-1mm}0.99&\hspace*{-2mm} -193.9 -- +193.9 & 9.7& 19.4& 11& 53\\
\hline
\end{tabular}
\end{center}
\end{minipage}
\end{table}
%
For more details on the observing and reduction procedures we refer to
Scharmer et al. (2008),
and Rouppe van der Voort et al. (2009).
In the latter publication, the 15-Jun-2008 data set has been analysed in the 
context of the connection between rapid blue-shifted excursions and type~II 
spicules.
Doppler shifts are determined as the shift of the line core with 
respect to the undisturbed line core wavelength of 854.2\,nm. 

\section{Analysis}
\label{sec:ana}

\paragraph{Atmospheric structure.} 
%
Traditionally, the prime diagnostic for the chromosphere has been the H$\alpha$ 
line. 
The IBIS observations by 
Cauzzi et al. (2008) and Vecchio et al. (2009)
 have demonstrated that the \ion{Ca}{II}~854.2~nm spectral 
line serves as an excellent alternative chromospheric diagnostic with the 
advantage of more straightforward interpretation as compared to H$\alpha$
%
(see e.g., Leenaarts et al. 2009). Cauzzi et al. (2008)
find that \ion{Ca}{II}~854.2~nm line core intensity maps reveal a fibrilar structure 
similar to what is well known from H$\alpha$ observations, and that 
fibrils cover 
a large fraction of the field of view even in more quiet regions. 
Next to the fibrils, which outline the magnetic field in the chromosphere, a
dynamic pattern produced by propagating shock waves 
(see e.g. Wedemeyer-B{\"o}hm et al. 2009)
is present. 
In contrast, our CRISP line core maps, which were obtained in coronal holes, 
exhibit only few fibrils, which cover only a limited fraction of the field of view. 
The magnetic field configuration in- and outside coronal holes is very 
different 
(see e.g. Cranmer 2009, and references therein), 
which has direct consequences for the morphology of the \caii\,854.2\,nm line core 
maps.
Similarly, the H$\alpha$ images directly taken after data set~2 show a
small number of fibrils. 

The scene of apparent magnetic quiescence is consistent with the atmosphere in the 
layers below, which show no pronounced magnetic network.
The FOV in the wide-band is characterised by granulation with bright points (BPs) 
in the intergranular lanes (see the upper row in Fig.~\ref{fig:swirl}). 
These BPs are known to be connected to magnetic field concentrations and 
are routinely used as proxy for the magnetic field.
The wide-band image essentially yields the continuum intensity and thus 
effectively maps the lower photosphere.
The map in the blue line wing at $\Delta \lambda = -92.1$\,pm  is clearly dominated 
by the reversed granulation pattern, which is formed in the middle photosphere 
(Leenaarts \& Wedemeyer-B{\"o}hm 2005).

\begin{figure}
\centering 
\includegraphics[width=7.8cm]{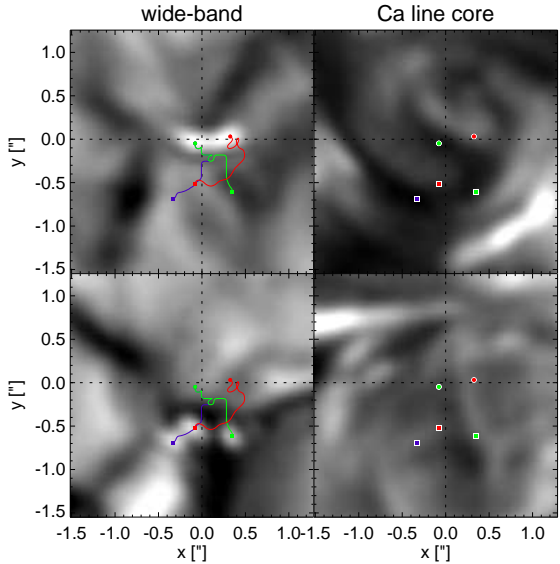}
\caption{Motion of photospheric bright points seen in the wide-band images (left 
column) at the position of a selected swirl (seen in the Ca line core images, right 
column). 
The two rows show two images with a separation of 10\,min. 
The initial positions of the BP tracks are marked with circles, the final ones with 
squares.  
}
\label{fig:bptrack}
\end{figure}

\paragraph{Swirls.}
\label{sec:swirl}
%
The line core time sequences show small regions with dark and bright rotating 
patches, which can consist of arcs, spiral arms, rings or ring fragments. 
In Fig.~\ref{fig:swirl} and the movie in the online material, we present a  
prominent example with a ring. 
The width of the ring is mostly $\sim 0\,\farcs2$, whereas the diameter is 
$\sim 2$\,\arcsec. 
The pattern can also be seen in the blue line wing at $\Delta \lambda = -53.3$\,pm 
and sometimes down to $\sim -70$\,pm, which is formed in the middle/upper photosphere.
At exactly the same position a bright ring-like counterpart appears in the red wing. 
The size of the pattern stays similar throughout this wavelength range. 
The line core is blue-shifted at positions in the ring, corresponding to upward 
(LOS) velocities typically of the order of $-$2 to $\sim -4$\,km$/$s. 
The swirl exhibits a small ring fragment with high Doppler shift at a 
position of [\mbox{$x\,\approx\,1\,\farcs0$}, \mbox{$y\,\approx\,0\,\farcs0$}] at 
$t = 0$\,s that appears to rotate clockwise (see Fig.~\ref{fig:swirl}).  
The Doppler shift of that line core feature is initially $-4.3$\,km$/$s and 
exceeds $\sim -5$\,km$/$s in the subsequent frames. 
The feature appears to have rotated by $\sim 90$\,degrees after roughly a minute. 
The centre of rotation is located in the central brightening visible in the line core maps, 
roughly coinciding with a BP group in the wide-band images.  
An estimate of the speed of the feature along its circular track yields 
approximate values of the order of $\sim 10$\,km$/$s. 
However, the feature evolves in time and cannot be tracked as reliably as wide-band BPs. 
The imprint of ring fragments is also seen in spectral line depth, being 
deepest where the line core intensity is lowest, and in the line width, 
measured at constant count levels. 
The latter is slightly increased at the rim of the ring. 

Another interesting aspect concerns the bright feature, which 
can be seen at a position of $[x \approx -1\,\farcs5, y \approx 0\,\farcs5]$ 
in the line core map at $t = 90$\,s (see Fig.~\ref{fig:swirl}).
It is located on a bright ring fragment just outside the dark ring and can be seen 
throughout the displayed sequence. 
The feature starts off rather faintly but increases in brightness as it 
follows the clockwise rotation. 
It is accompanied by a strong Doppler shift which grows in time 
to up to -6.9\,km$/$s before it quickly fades away after $t = 90$\,s. 
The line wing and wide-band maps show nothing special in these moments so that this 
event is likely to be confined to the layers above the photosphere. 
More of these features appear during the swirl event.

We interpret the blue-shift and the clockwise rotation as plasma spiraling upwards 
in a funnel-like magnetic field structure with a diameter of $\sim 2$\,\arcsec.  
It may be due to propagating wave fronts, which are guided by the magnetic field.  
The short-lived bright features with high Doppler shift may be explained as 
plasma which is accelerated and then ejected upwards and away from the swirl at 
large velocity.

\paragraph{Statistics.}  
%
Swirls are frequently observed in both our coronal hole data sets. 
For a measure of the swirl statistics, we performed a thorough inspection of 
dataset~1. 
We found 10 clear examples, including the one discussed above. 
In addition, we found 10 cases which feature (minor) swirl action, arcs, spirals 
and/or ring fragments for a short time span but not as strikingly and lastingly 
as in the clear detections mentioned before. 
Finally, there are another 16 potential but less reliable detections. 
Unambiguously  discerning a faint swirl imprint requires high data quality over a 
sufficiently long period. 
That can be hampered by variation in the observation conditions and resulting low 
contrast but also by the presence of fibrils.  
It is therefore difficult to determine reliable event durations. 
We see swirls that can only be identified for a few consecutive time steps and
other swirls that last for minutes and/or reappear at the same location later in 
the time series. 
In the region presented in Fig.~\ref{fig:swirl}, swirling motions are detected
for about 12\,min of which the first 8\,min are very clear. 
A conservative estimate of the frequency of swirl event occurrence yields a value of 
0.24\,swirls\,min$^{-1}$\,arcmin$^{-2}$ based on the clear detections. 
It increases to 0.48\,swirls\,min$^{-1}$\,arcmin$^{-2}$ when taking into account 
the potential detections, too.

\begin{figure}
\centering 
\includegraphics[width=6.5cm]{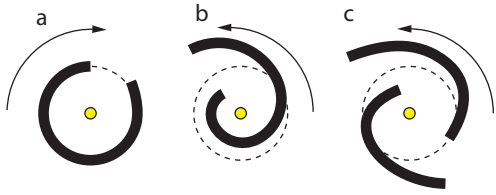}
\caption{Schematic structure of swirl events. The radius of the dotted circle is 
typically of the order of 1\arcsec.}
\label{fig:swirlsketch}
\end{figure}

The details of the swirl pattern in the atmosphere above varies from case to case.  
The rings in the line core maps have radii of typically 1\,arcsec, although 
there are examples with radii down to 0\,\farcs4 - 0\,\farcs6 and up to 1\,\farcs5. 
The line core rings can be seen during most of the event, while the Doppler shift 
signature can fade away and reappear again. 
We also find some cases for which spiral arms are visible instead of or in addition 
to a roughly concentric ring (see Fig.~\ref{fig:swirlsketch}b-c). 
The inner parts of such spiral arms are found 0\,\farcs4 to 1\,\farcs0 away from 
the BP, while the outer tip, which is leading the rotation, can be traced up to 
a distance of $\sim 2$\arcsec.  
Dopplergrams show that the blue-shift of the line core tends to increase along 
the dark spiral arms outwards and can reach upward velocities of up to 
$\sim 7$\,km$/$s. 
In some examples it appears as if plasma is accelerated and 
then ejected away from the central BP group. 
 
Based on the swirls identified, there does not seem to be a preferred direction of 
rotation. 
Also the location within the observed coronal hole appears to be random.

\paragraph{Photospheric bright point groups.} 
%
The line wing and wide-band images, which map the middle and low photosphere,
show small groups of photospheric BPs at the locations of all detected swirls. 
These photospheric BPs move vigorously along the intergranular lanes and 
the brights rims of the reversed granulation pattern and  below, indicating that 
they are buffeted by the lateral granular flows. 
In Fig.~\ref{fig:bptrack} we show the movement of the photospheric BPs for the 
example presented in Fig.~\ref{fig:swirl}. 
A bright elongated structure is seen in the initial wide-band image (upper row) 
at the location of the swirl. 
The brightening appears to consist of at least 2 (unresolved) BPs. 
By using a feature tracking algorithm, we find that the initial bright structure 
breaks apart into at least three individual BPs with separations of only a few 
tenths of an arcsecond. 
The BPs move with average speeds between 1.5 and 2.0\,km/s for the example in 
Fig.~\ref{fig:bptrack}. 
The relative motion pattern is subject to the local granular flow field and thus varies 
from case to case. 
Generally, BPs related to swirls are found to have lateral speeds between 0.7 and 
2.0\,km/s, which is consistent with the expected advection of the BPs with the granular 
flow field. 
In Fig.~\ref{fig:bptrack}, two BPs seem to cross their paths, which has consequences 
for the magnetic field topology in the chromosphere above. 
The other swirls also show close encounters of BPs. 
Although it is sometimes difficult to resolve all individual components, an apparently 
chaotic relative motion of BPs seems to be common below swirl regions.     

\section{Discussion and conclusions}
\label{sec:discus}

At first glance, the swirls in the coronal hole are reminiscent of the 
convectively driven vortex flows reported by Bonet et al. (2008).
In a high-resolution SST G-band filtergram time sequence, they found examples of 
groups of BPs displaying clear ``whirlpool'' motion. 
Bonet et al. 
 associate the photospheric whirlpools with vigorous 
downdrafts at the vertices of intergranular lanes that have been predicted by 
convection models. 
The G-band BPs act as flow tracers which follow spiral trajectories when they 
are engulfed by the downdrafts.
The size of these whirlpools is less than an arcsecond. 
Contrastingly, the photospheric BPs associated with the chromospheric swirls in our data do not 
follow clear spiral trajectories. 
It is therefore unclear if photospheric whirlpools and chromospheric swirls are connected. 

Our wide-band images usually show a small number of photospheric BPs that move in 
relation to each other, buffeted by the lateral granular flows at the top of the 
convection zone. 
Being confined to the intergranular lanes, the BPs can eventually come close and even 
move past each other very closely. 
Our interpretation is that these close encounters result in a deformation and twist of the 
magnetic field continuing above the BPs. 
While tending to unwind the build-up stresses, gas may spiral upwards, producing the 
observed intensity signature of Doppler-shifted ring fragments.  

Such a process may be impeded by the presence of a stronger magnetic field 
component in the chromosphere in the form of a ``magnetic canopy'' 
(e.g., Schrijver \& Title 2003) 
as it exists above quiet Sun regions outside coronal holes. 
Swirls are therefore expected to be significantly less frequent in quiet 
Sun regions outside coronal holes, which have a high fibril coverage. 
Quantifying this assumption is unfortunately difficult as the detection of swirls in 
Ca line core maps is complicated by the presence of fibrils. 
Events ongoing in the layer just below the fibrils could be obscured by the horizontal 
component of the magnetic field (Vecchio et al. 2009).

The exact cause for the swirl phenomenon remains to be found. 
The magnetic fields, which are supposedly responsible for forcing the rotating 
plasma or propagating waves on ring-like or spiral trajectories, are not directly 
detected in our CRISP observations. 
However, all our detected swirls show a small group of BPs in the centre, 
implying a causal connection. 

We interpret these chromospheric swirl motions and associated BP motions as a 
direct indication of upper-atmospheric magnetic field twisting and braiding as a 
result of convective buffeting of magnetic footpoints. 
This mechanism is one of the prime candidates for coronal heating 
(Parker 1988).

\begin{acknowledgements}
This study made use of the CRISPEX data explorer programmed by 
Gregal Vissers. 
The Swedish 1-m Solar Telescope is operated on the island of La Palma
by the Institute for Solar Physics of the Royal Swedish Academy of
Sciences in the Spanish Observatorio del Roque de los Muchachos of the
Instituto de Astrof{\'\i}sica de Canarias.
SWB acknowledges support through a Marie Curie Intra-European Fellowship of the 
European Commission (FP6-2005-Mobility-5, Proposal No. 
042049).   
This research was supported by the Research Council of Norway through
grant 170935/V30.
\end{acknowledgements}

\bibliographystyle{aa} 
\section*{References}
\tiny

%
\noindent
{Bonet}, J.~A., {M{\'a}rquez}, I., {S{\'a}nchez Almeida}, J., {Cabello}, I., \&
  {Domingo}, V. 2008, \apjl, 687, L131

\noindent
{Brandt}, P.~N., {Scharmer}, G.~B., {Ferguson}, S., {Shine}, R.~A., \&
  {Tarbell}, T.~D. 1988, \nat, 335, 238

\noindent
{Brown}, D.~S., {Nightingale}, R.~W., {Alexander}, D., {et~al.} 2003, \solphys,
  216, 79

\noindent
{Cauzzi}, G., {Reardon}, K.~P., {Uitenbroek}, H., {et~al.} 2008, \aap, 480, 515

\noindent
{Cavallini}, F. 2006, \solphys, 236, 415

\noindent
{Cranmer}, S.~R. 2009, ArXiv e-prints 0909.2847

\noindent
{Judge}, P. 2006, in Astronomical Society of the Pacific Conference Series,
  Vol. 354, Solar MHD Theory and Observations: A High Spatial Resolution
  Perspective, ed. J.~{Leibacher}, R.~F. {Stein}, \& H.~{Uitenbroek}, 259

\noindent
{Leenaarts}, J., {Carlsson}, M., {Hansteen}, V., \& {Rouppe van der Voort}, L.
  2009, \apjl, 694, L128

\noindent
{Leenaarts}, J. \& {Wedemeyer-B{\"o}hm}, S. 2005, \aap, 431, 687

\noindent
{Parker}, E.~N. 1983, \apj, 264, 642

\noindent
{Parker}, E.~N. 1988, \apj, 330, 474

\noindent
{Puschmann}, K.~G., {Kneer}, F., {Seelemann}, T., \& {Wittmann}, A.~D. 2006,
  \aap, 451, 1151

\noindent
{Rouppe van der Voort}, L., {Leenaarts}, J., {de Pontieu}, B., {Carlsson}, M.,
  \& {Vissers}, G. 2009, ArXiv e-prints 0909.2115

\noindent
{Rutten}, R.~J. 2006, in Astronomical Society of the Pacific Conference Series,
  Vol. 354, Solar MHD Theory and Observations: A High Spatial Resolution
  Perspective, ed. J.~{Leibacher}, R.~F. {Stein}, \& H.~{Uitenbroek}, 276

\noindent
{Rutten}, R.~J. 2007, in Astronomical Society of the Pacific Conference Series,
  Vol. 368, The Physics of Chromospheric Plasmas, ed. P.~{Heinzel},
  I.~{Dorotovi{\v c}}, \& R.~J. {Rutten}, 27

\noindent
{Scharmer}, G.~B., {Bjelksjo}, K., {Korhonen}, T.~K., {Lindberg}, B., \&
  {Petterson}, B. 2003, in Society of Photo-Optical Instrumentation Engineers
  (SPIE) Conference Series, ed. S.~L. {Keil} \& S.~V. {Avakyan}, Vol. 4853, 341

\noindent
{Scharmer}, G.~B., {Narayan}, G., {Hillberg}, T., {et~al.} 2008, \apjl, 689,
  L69

\noindent
{Schrijver}, C.~J. 2001, in Astronomical Society of the Pacific Conference
  Series, Vol. 223, 11th Cambridge Workshop on Cool Stars, Stellar Systems and
  the Sun, ed. R.~J. {Garcia Lopez}, R.~{Rebolo}, \& M.~R. {Zapaterio Osorio},
  131

\noindent
{Schrijver}, C.~J. \& {Title}, A.~M. 2003, 597, L165

\noindent
{van Noort}, M., {Rouppe van der Voort}, L., \& {L{\"o}fdahl}, M.~G. 2005,
  \solphys, 228, 191

\noindent
{Vecchio}, A., {Cauzzi}, G., \& {Reardon}, K.~P. 2009, \aap, 494, 269

\noindent
{Wedemeyer-B{\"o}hm}, S., {Lagg}, A., \& {Nordlund}, {\AA}. 2009, Space Science
  Reviews, 144, 317
%
\end{document}